\documentclass[conference, 10pt]{IEEEtran}
\IEEEoverridecommandlockouts
\usepackage{cite}
\usepackage{amsmath,amssymb,amsfonts}
\usepackage{algorithmic}
\usepackage{graphicx}
\usepackage{textcomp}
\usepackage{xcolor}
\def\BibTeX{{\rm B\kern-.05em{\sc i\kern-.025em b}\kern-.08em
    T\kern-.1667em\lower.7ex\hbox{E}\kern-.125emX}}
\begin{document}
\onecolumn
\textbf{Disclaimer}

\vspace{0.2in} 

{Paper accepted for presentation in IEEE SPAWC 2021 - 22nd IEEE International Workshop on Signal Processing Advances in Wireless Communications.}
\vspace{0.5in}

{\textcopyright \ 2021 IEEE. Personal use of this material is permitted. Permission from IEEE must be obtained for all other uses, in any current or future media, including reprinting/republishing this material for advertising or promotional purposes, creating new collective works, for resale or redistribution to servers or lists, or reuse of any copyrighted component of this work in other works.}

\twocolumn
\newpage 

\title{Optimal Joint Beamforming and Power Control in Cell-Free Massive MIMO Downlink\\
\thanks{This work was supported by H2020 Marie Skłodowska-Curie Actions (MSCA) Individual Fellowships (IF) IUCCF, grant 844253.
}
}

\author{\IEEEauthorblockN{Mohamed Elwekeil$^{1,2}$,  Alessio Zappone$^{1,3}$, and Stefano Buzzi$^{1,3}$}
\IEEEauthorblockA{$^{1}$\textit{Department of Electrical and Information Engineering, University of Cassino and Southern Lazio, Cassino, Italy.} \\
(email:\{mohamed.elwekeil, alessio.zappone, s.buzzi\}@unicas.it).\\
$^{2}$\textit{Department of Electronics and Electrical Communications Engineering, Faculty of Electronic Engineering,} \\ \textit{Menoufia University, Menouf 32952, Egypt.}  (email: mohamed.elwekeil@el-eng.menofia.edu.eg),\\ $^{3}$ \textit{Consorzio Nazionale Interuniversitario per le Telecomunicazioni, Parma, Italy}.
}
}

\maketitle

\begin{abstract}
In this paper, a novel optimization model for joint beamforming and power control in the downlink (DL) of a cell-free massive MIMO (CFmMIMO) system is presented. The objective of the proposed optimization model is to minimize the maximum user interference while satisfying quality of service (QoS) constraints and power consumption limits. The proposed min-max optimization model is formulated as a mixed-integer nonlinear program, that is directly tractable. Numerical results show that the proposed joint beamforming and power control scheme is effective and outperforms competing schemes in terms of data rate, power consumption, and energy efficiency. 
 
\end{abstract}

\begin{IEEEkeywords}
Cell-free massive MIMO, beamforming, power control, min-max optimization, energy efficiency
\end{IEEEkeywords}

\section{Introduction}
%
 
Cell-free massive MIMO (CFmMIMO) has attracted the attention of the research community as one of the most promising technologies that can be used in beyond fifth-generation (5G) networks. The concept of CFmMIMO is presented in \cite{ngo2015cell}, where the antennas are distributed, in the form of access points (APs), in the service area instead of being collocated at a cell center. CFmMIMO is compared to small cells in \cite{ngo2017cell}, wherein the authors concluded that CFmMIMO outperforms small cells in terms of throughput. In the generic CFmMIMO, APs in the service area, even those that are far from the end-user, should participate in providing the service for the end-user. This is considered a waste of energy as the APs far from the end-user will have poor signal to interference plus noise ratios (SINRs). 

Thus, a user-centric (UC) approach for CFmMIMO is proposed in \cite{buzzi2017cell}, where service is provided to the end-user through a subset of the APs, the closest to it. The energy efficiency of CFmMIMO has been investigated in \cite{ngo2017total}, where the authors proposed power control and AP selection algorithms to improve the energy efficiency of the CFmMIMO system. However, most of these works assume the utilization of conjugate beamforming and formulate the resource allocation problem in terms of the throughput, which is usually non-convex. Thus, these works resort to approximations and needs to call the optimization solver multiple times leading to sub-optimal solutions. 
 
In this paper, we are dealing with the joint beamforming and power control problem in a UC-CFmMIMO system that serves multiple users by presenting a novel formulation. Specifically, we propose a new optimization model with the objective to minimize the maximum user interference and satisfy the quality of service (QoS) and energy consumption constraints. To the best of the authors' knowledge, this work is the first to consider this min-max interference formulation and presents a new framework for tackling the resource allocation problem is wireless communications. We managed to represent the proposed min-max optimization model in a form that can be directly tackled using the cvx optimization toolbox \cite{cvx}. 
 
\section{System Model} \label{sec:system}

\begin{figure}[!t]
\begin{center}
\includegraphics[width =0.75\columnwidth]{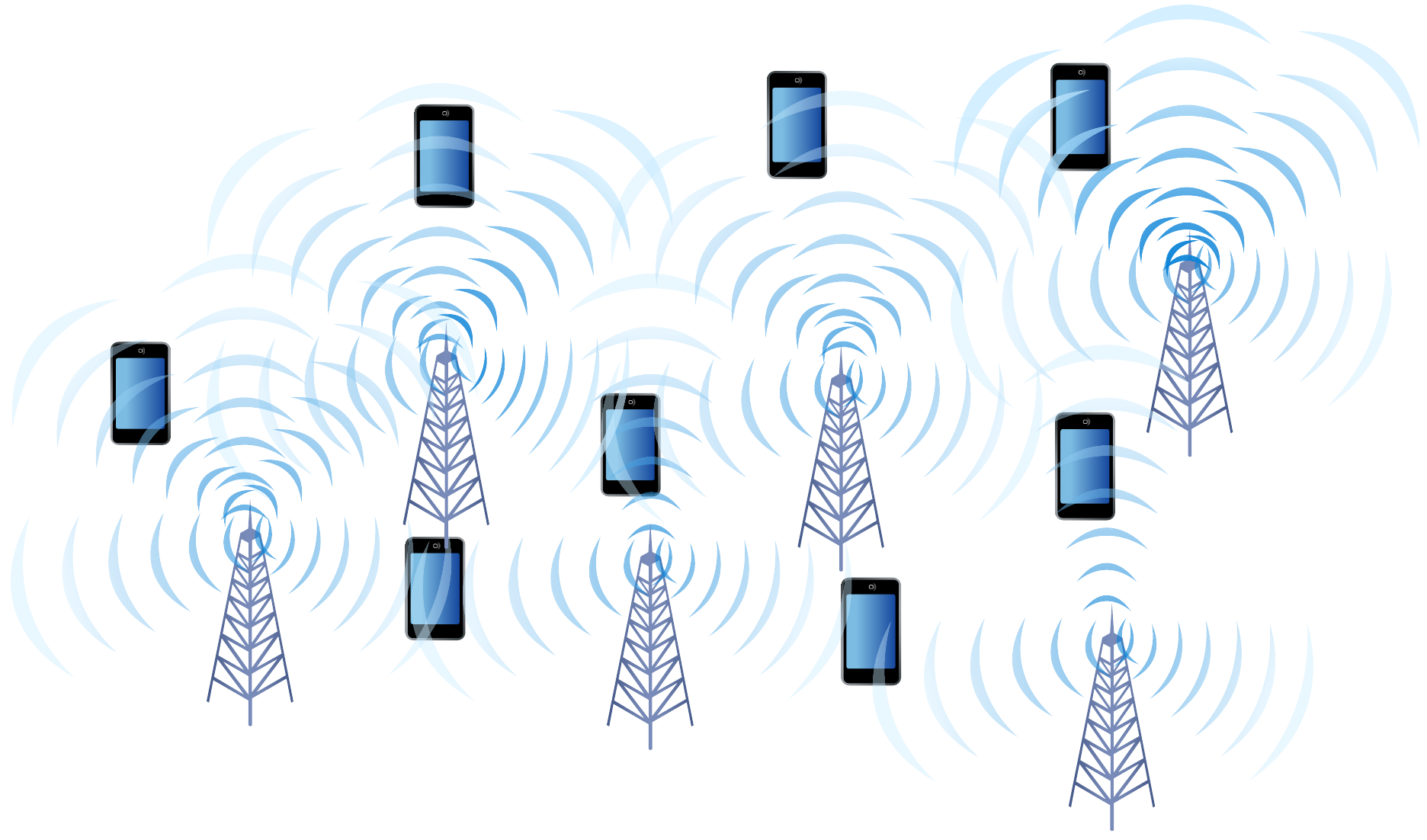}
\caption{The considered cell-free massive MIMO system.}
\label{fig:system model}
\end{center}
\end{figure}

The considered system is depicted in Fig. \ref{fig:system model}. Particularly, we consider a CFmMIMO network that consists of a set of access points $\mathbb{A}=\{AP_1,AP_2,..,AP_a,..,AP_{N_A}\}$, where $N_A=\#\{\mathbb{A}\}$ is the cardinality of the set $\mathbb{A}$, which is the number of available APs in the network. We assume that all APs are connected via ideal backhaul to a central processing unit (CPU). Let $\mathbb{U}=\{U_1,U_2,..,U_u..,U_{N_U}\}$ denotes the set of the users , where $N_U=\#\{\mathbb{U}\}$ is the cardinality of the set $\mathbb{U}$, which is the number of users. 

Every AP in the network employs $M_{AP}$ antennas, while every user utilizes only a single antenna. We follow the time division duplex (TDD) transmission framework typically adopted in the literature \cite{ngo2015cell, ngo2017cell, buzzi2017cell, ngo2017total, 8952782}. Specifically, the coherence interval, $\tau_c$, (also referred to as time slot) is employed for uplink  (UL) channel estimation during sub-interval $\tau_p<\tau_c$, downlink (DL) data transmission during sub-interval $\tau_d<\tau_c$, and UL data transmission during sub-interval $\tau_u<\tau_c$. We assume perfect channel state information at the AP, where the actual channel coefficients are known at each AP. 

We adopt a user-AP association approach wherein each user $u$ is served by only a subset of the existing APs in the network based on the large scale fading coefficients \cite{ngo2017total,8952782}. Thus, each user $u$ is served by a subset of the available APs, specifically, $\mathbb{A}_{u}=\{AP_{1,u},AP_{2,u},..,AP_{N_{A,u}}\}$, where $N_{A,u}$ is the number of APs serving the user $u$. Specifically, user $u$ is served by $N_{A,u}$ APs that have the best channel condition to this user. We hereby, denote by $\mathbb{U}_a=\{U_{1,,a}, U_{2,a},...,U_{N_{U,a}}\} $ the set of users served by AP $a$. 
The channel vector from user $u$ to AP $a$ in the time slot $n$ is denoted by $\boldsymbol{h}_{u,a}[n]\in \mathbb{C}^{M_{AP}\times 1}$. For every user, we consider that the small scale fading is Rayleigh. 
Thus, the channel vector from user $u$ to the AP $a$ in the time slot $n$, $\boldsymbol{h}_{u,a} [n]$ is expressed as
\begin{equation}
\label{eq:eq-1_channel}
 \begin{array}{ll}
\boldsymbol{h}_{u,a}[n]& = \sqrt{\beta_{u,a}[n]}\boldsymbol{g}_{u,a}[n],
\end{array}
\end{equation}
where $\beta_{u,a}[n] \in \mathbb{R}$ is the large scale fading coefficient which includes the effects of both path loss and shadowing between user $u$ and AP $a$, and $\boldsymbol{g}_{u,a}[n]\in \mathbb{C}^{M_{AP}\times 1}$ includes the Rayleigh fading coefficients between user $u$ and AP $a$, i.e., $\boldsymbol{g}_{u,a}[n] \sim  \mathcal{CN}(\boldsymbol{0}, \boldsymbol{I}_{M_{AP}})$, with $\boldsymbol{I}_{M_{AP}}$ is the ${M_{AP}} \times {M_{AP}}$ identity matrix.

In this work, we concentrate on the DL of the considered scenario where the APs in the network are transmitting to multiple users. The signal transmitted from AP $a$ in the time slot $n$, $\boldsymbol{x}_a[n]$ is given by
\begin{equation}
\label{eq:eq-1}
\begin{array}{ll}
\boldsymbol{x}_{a}[n]& = \displaystyle \sum_{u \in\mathbb{U}_a}\boldsymbol{f}_{a,u}[n]s_u^{dl}[n],
\end{array}
\end{equation}
where $\boldsymbol{f}_{a,u}[n]\in \mathbb{C}^{M_{AP}\times 1}$ is the DL beamforing vector from AP $a$ to user $u$ in the time slot $n$, and $s_u^{dl}[n]$ is the DL data symbol to be transmitted for user $u$ in the time slot $n$. Note that $ \left \| \boldsymbol{f}_{a,u}[n] \right \|^2=\alpha_{a,u}[n]$ is the corresponding DL power control coefficient from AP $a$ to user $u$. 

The received signal at user $u$ in the time slot $n$, $y_u[n]$, is given by 
\begin{equation}
\label{eq:eq-3}
\begin{array}{ll}
y_u[n]& = \displaystyle \sum_{a \in\mathbb{A}}\boldsymbol{h}_{u,a}^*[n]\boldsymbol{x}_{a}[n]+w_u[n]
\\
  & = \displaystyle \sum_{a \in\mathbb{A}} \displaystyle \sum_{i \in\mathbb{U}_a}\boldsymbol{h}_{u,a}^*[n]\boldsymbol{f}_{a,i}[n]s_i^{dl}[n]+w_u[n],
\end{array}
\end{equation} 
where $w_u[n]\sim \mathcal{CN}(0,\sigma_w^2)$ is the additive white Gaussian noise (AWGN) at user $u$ in the time slot $n$, and $\sigma_w^2$ is the noise variance. Equation (\ref{eq:eq-3}) can be represented in an equivalent form such that the desired signal and the interference terms at user $u$ are separable as follows
\begin{equation}
\label{eq:eq-4}
\begin{array}{ll}
y_u[n]& = \displaystyle \sum_{i \in\mathbb{U}} \displaystyle \sum_{a \in\mathbb{A}_i}\boldsymbol{h}_{u,a}^*[n]\boldsymbol{f}_{a,i}[n]s_i^{dl}[n]+w_u[n]
\\
& = \displaystyle \sum_{a \in\mathbb{A}_u}\boldsymbol{h}_{u,a}^*[n]\boldsymbol{f}_{a,u}[n]s_u^{dl}[n]
\\
& + \displaystyle \sum_{i \in\mathbb{U}\setminus u} \displaystyle \sum_{a \in\mathbb{A}_i}\boldsymbol{h}_{u,a}^*[n]\boldsymbol{f}_{a,i}[n]s_i^{dl}[n]+w_u[n],
\end{array}
\end{equation} 
where the first term is the desired signal term, the second term is the interference term, and the third term is the AWGN term. 

\section{Proposed Joint Beamforming and Power Control} \label{sec:Proposed}
Usually, the wireless resource allocations are formulated as optimization models with objectives based on the data rate (e.g. maximize the total network sum rate or maximizing the minimum user's data rate). However, the resultant optimization model is usually non-convex and is handled by multiple calls to a solver, which leads to longer convergence times. In this paper, we resort to a new paradigm in formulating the resource allocation problem, where our objective is to minimize the maximum interference at the users' level while satisfying constraints on the desired received signal and the power budget consumption. Although this work applies this framework to solve the joint beamforming and power control problem, the same framework can be employed to tackle other resource allocation problems. 

From equation (\ref{eq:eq-4}), it is clear that the desired signal power received at user $u$ in the time slot $n$, $P_u^S[n]$, is expressed by
\begin{equation}
\label{eq:eq-5}
\begin{array}{ll}
P_u^S[n] & = \left | \displaystyle \sum_{a \in\mathbb{A}_u}\boldsymbol{h}_{u,a}^*[n]\boldsymbol{f}_{a,u}[n]\right |^2,
\end{array}
\end{equation} 
and the interference power at user $u$ in the time slot $n$, $P_u^I[n]$, is given by
\begin{equation}
\label{eq:eq-6}
\begin{array}{ll}
P_u^I[n] & = \displaystyle \sum_{i \in\mathbb{U}\setminus u} \displaystyle  \left | \sum_{a \in\mathbb{A}_i}\boldsymbol{h}_{u,a}^*[n]\boldsymbol{f}_{a,i}[n]\right |^2.
\end{array}
\end{equation}
Thus, the corresponding DL SINR of user $u$ in the time slot $n$, $\gamma_u^{dl}[n]$, is given by
\begin{equation}
\label{eq:eq-7}
\begin{array}{ll}
\gamma_u^{dl}[n]& = \displaystyle \frac{\left | \displaystyle \sum_{a \in\mathbb{A}_u}\boldsymbol{h}_{u,a}^*[n]\boldsymbol{f}_{a,u}[n]\right |^2}{ \displaystyle \sum_{i \in\mathbb{U}\setminus u} \left | \displaystyle \sum_{a \in\mathbb{A}_i}\boldsymbol{h}_{u,a}^*[n]\boldsymbol{f}_{a,i}[n]\right |^2+\sigma_w^2}.
\end{array}
\end{equation}
Hence, the DL data rate of user $u$ is given by
\begin{equation}
\label{eq:eq-7_2}
\begin{array}{ll}
R_u^{dl}[n]=\displaystyle \frac{\tau_d}{\tau_c}B \log_2(1+\gamma_u^{dl}[n]),
\end{array}
\end{equation}
where $B$ is the channel bandwidth.

We propose a new formulation for the joint beamforming and power control problem in CFmMIMO network, where the objective is to minimize the worst user's interference. The proposed optimization problem can be expressed as follows
\begin{equation}
\label{eq:eq-10_2}
\begin{array}{l}
\underset{\boldsymbol{f}_{a,u}[n], \forall u,a}{\min}  \underset{u}{\max} \ \displaystyle \sum_{i \in\mathbb{U}\setminus u} \displaystyle  \left | \sum_{a \in\mathbb{A}_i}\boldsymbol{h}_{u,a}^*[n]\boldsymbol{f}_{a,i}[n]\right |^2
\\
\textrm{s.t.} \\
\rho_u \leq \left | \displaystyle \sum_{a \in\mathbb{A}_u}\boldsymbol{h}_{u,a}^*[n]\boldsymbol{f}_{a,u}[n]\right |^2 \leq \mu_u, \ \forall u\in \mathbb{U},
\\
 \displaystyle \sum_{u \in \mathbb{U}_a}\left \| \boldsymbol{f}_{a,u}[n]\right \|^2 \leq \eta_a, \ \forall a \in \mathbb{A}. 
\end{array}
\end{equation}
The objective of the optimization model (\ref{eq:eq-10_2}) is to find the set of beamforming vectors $\boldsymbol{f}_{a,u}[n], \forall u,a$ such that the worst user's interference is minimized. The interference at user $u$ is given as in equation (\ref{eq:eq-6}). The first constraint of the optimization model (\ref{eq:eq-10_2}) sets a lower bound $\rho_u$ and an upper bound $\mu_u$ on the received desired signal $P_u^S[n]$. The lower bound $\rho_u$ is employed to ensure that at least a certain power level for the desired signal is received at the user and thus a corresponding minimum QoS  can be guaranteed. On the other hand, the upper bound $\mu_u$ is imposed to prevent users that have a very good channel condition from absorbing the whole available resources that might impact other users with inferior channel conditions. The second constraint of the optimization model (\ref{eq:eq-10_2}) assures that the power transmitted by each AP satisfies a maximum power budget, $\eta_a$.

The objective and the first constraint of optimization model (\ref{eq:eq-10_2}) can be reformulated so that the model (\ref{eq:eq-10_2}) is rewritten as 
\begin{equation}
\label{eq:eq-10_5}
\begin{array}{l}
\underset{\boldsymbol{f}_{a,u}[n], \forall u,a}{\min}  \underset{u}{\max} \ \displaystyle \sum_{i \in\mathbb{U}\setminus u} \displaystyle  \left | \sum_{a \in\mathbb{A}_i}\boldsymbol{h}_{u,a}^*[n]\boldsymbol{f}_{a,i}[n]\right |
\\
\textrm{s.t.} \\
\sqrt{\rho_u} \leq \left | \displaystyle \sum_{a \in\mathbb{A}_u}\boldsymbol{h}_{u,a}^*[n]\boldsymbol{f}_{a,u}[n]\right | \leq \sqrt{\mu_u}, \  \forall u\in \mathbb{U},
\\
\displaystyle \sum_{u \in \mathbb{U}_a}\left \| \boldsymbol{f}_{a,u}[n]\right \|^2 \leq \eta_a, \ \forall a \in \mathbb{A}. 
\end{array}
\end{equation}
The min-max objective of the optimization model (\ref{eq:eq-10_5}) can be represented in a linear form by introducing a new variable $z$, that takes the role of the maximum part as follows \cite{bisschop2006aimms}
\begin{equation}
\label{eq:eq-11}
\begin{array}{ll}
z& = \underset{u}{\max} \displaystyle \sum_{i \in\mathbb{U}\setminus u} \displaystyle  \left | \sum_{a \in\mathbb{A}_i}\boldsymbol{h}_{u,a}^*[n]\boldsymbol{f}_{a,i}[n]\right |.
\end{array}
\end{equation}
Consequently, the following new constraint should be imposed
\begin{equation}
\label{eq:eq-12}
\begin{array}{ll}
z& \geq \displaystyle \sum_{i \in\mathbb{U}\setminus u} \displaystyle  \left | \sum_{a \in\mathbb{A}_i}\boldsymbol{h}_{u,a}^*[n]\boldsymbol{f}_{a,i}[n]\right |, \ \forall u \in \mathbb{U}.
\end{array} 
\end{equation}

At this point, when $z$ is minimized, the constraint (\ref{eq:eq-12}) assures that $z$ should be greater than or equal to $\displaystyle \sum_{i \in\mathbb{U}\setminus u} \displaystyle  \left | \sum_{a \in\mathbb{A}_i}\boldsymbol{h}_{u,a}^*[n]\boldsymbol{f}_{a,i}[n]\right |, \forall u \in \mathbb{U}$. Simultaneously, given the fact that $z$ is minimized, the optimal value of $z$ will not be greater than the maximum value of $\displaystyle \sum_{i \in\mathbb{U}\setminus u} \displaystyle  \left | \sum_{a \in\mathbb{A}_i}\boldsymbol{h}_{u,a}^*[n]\boldsymbol{f}_{a,i}[n]\right |$. Thus, the optimal value of $z$ should be minimized and will exactly equal the maximum value of $\displaystyle \sum_{i \in\mathbb{U}\setminus u} \displaystyle  \left | \sum_{a \in\mathbb{A}_i}\boldsymbol{h}_{u,a}^*[n]\boldsymbol{f}_{a,i}[n]\right |$ \cite{bisschop2006aimms}. Therefore, the optimization model (\ref{eq:eq-10_5}) can be written in an equivalent form (\ref{eq:eq-13})
\begin{equation}
\label{eq:eq-13}
\begin{array}{l}
\underset{\boldsymbol{f}_{a,u}[n], \forall u,a}{\min} ~ z
\\
\textrm{s.t.} \\
 \displaystyle \sum_{i \in\mathbb{U}\setminus u} \displaystyle  \left | \sum_{a \in\mathbb{A}_i}\boldsymbol{h}_{u,a}^*[n]\boldsymbol{f}_{a,i}[n]\right |-z \leq 0, \ \forall u\in \mathbb{U},
\\  \sqrt{\rho_u} \leq \left | \displaystyle \sum_{a \in\mathbb{A}_u}\boldsymbol{h}_{u,a}^*[n]\boldsymbol{f}_{a,u}[n]\right | \leq \sqrt{\mu_u}, \  \forall u\in \mathbb{U},
\\
 \displaystyle \sum_{u \in \mathbb{U}_a}\left \| \boldsymbol{f}_{a,u}[n]\right \|^2 \leq \eta_a, \ \forall a \in \mathbb{A}. 
\end{array}
\end{equation}
It is noteworthy that the objective of optimization model (\ref{eq:eq-13}) along with its first constraint play the same role as the objective of the optimization model (\ref{eq:eq-10_5}). Besides, the remaining constraints of the optimization model (\ref{eq:eq-13}) are exactly the same as the constraints of the optimization model (\ref{eq:eq-10_5}). 

By introducing slack variables $V_u^{S,+}$ and $V_u^{S,-}$, it is possible to remove the modulus operator in the second inequality constraint in (\ref{eq:eq-13}), and to reformulate it as the equality constraint\footnote{Note that having complex quantities poses no mathematical issue since the constraint is now cast in equality form.} $\sum_{a \in\mathbb{A}_u}\boldsymbol{h}_{u,a}^*[n]\boldsymbol{f}_{a,u}[n] = V_u^{S,+} - V_u^{S,-}$, where $ V_u^{S,+} \geq 0$ and $ V_u^{S,-} \geq 0$ are slack variables that take the positive and negative values of $\sum_{a \in\mathbb{A}_u}\boldsymbol{h}_{u,a}^*[n]\boldsymbol{f}_{a,u}[n]$, respectively. It can be seen that, at the optimum, at least one of the slack variables $ V_u^{S,+}$ and $ V_u^{S,-}$ must equal zero. Thus, an either or constraint is introduced as in \cite{bisschop2006aimms, elwekeil2012optimal}  
\begin{equation}
\label{eq:eq-14}
\begin{array}{l}
  V_u^{S,+} \leq \Delta \varphi_u, \ \forall u\in \mathbb{U},
\\
  V_u^{S,-} \leq \Delta (1-\varphi_u), \ \forall u\in \mathbb{U},
\end{array}
\end{equation}
where $\varphi_u$ is a binary slack variable and $\Delta $ is a coefficient that should be large enough to have satisfactory accuracy. 
 Specifically, $ \displaystyle \sum_{a \in\mathbb{A}_u}\boldsymbol{h}_{u,a}^*[n]\boldsymbol{f}_{a,u}[n] = V_u^{S,+}$ and $\left | \displaystyle \sum_{a \in\mathbb{A}_u}\boldsymbol{h}_{u,a}^*[n]\boldsymbol{f}_{a,u}[n]\right | = V_u^{S,+}$ if $ \varphi_u =1$; while $ \displaystyle \sum_{a \in\mathbb{A}_u}\boldsymbol{h}_{u,a}^*[n]\boldsymbol{f}_{a,u}[n] = - V_u^{S,-}$ and $\left | \displaystyle \sum_{a \in\mathbb{A}_u}\boldsymbol{h}_{u,a}^*[n]\boldsymbol{f}_{a,u}[n]\right | = V_u^{S,-}$ if $ \varphi_u =0$. 

Accordingly, the second constraint in the optimization model (\ref{eq:eq-13}) can be represented by the  set of constraints
\begin{equation}
\label{eq:eq-15}
\begin{array}{l}
\sqrt{\rho_u} \leq V_u^{S,+} + V_u^{S,-} \leq \sqrt{\mu_u}, \ \forall u\in \mathbb{U},
\\
\displaystyle \sum_{a \in\mathbb{A}_u}\boldsymbol{h}_{u,a}^*[n]\boldsymbol{f}_{a,u}[n] - V_u^{S,+} + V_u^{S,-} = 0, \ \forall u\in \mathbb{U},
\\
V_u^{S,+} \leq \Delta \varphi_u, \ \forall u\in \mathbb{U},
\\
V_u^{S,-} \leq \Delta (1-\varphi_u), \ \forall u\in \mathbb{U},
\\
V_u^{S,+} \geq 0, \  \forall u\in \mathbb{U},
\\ V_u^{S,-} \geq 0, \ \forall u\in \mathbb{U},
\\ \varphi_u \in \{0,1\}, \ \forall u\in \mathbb{U}.  
\end{array}
\end{equation}

Now, following the approach in \cite{bisschop2006aimms}, it is convenient to represent the range constraint in (\ref{eq:eq-15}) by introducing slack variables $\nu_u$ as follows 
\begin{equation}
\label{eq:eq-17}
\begin{array}{l}
V_u^{S,+} + V_u^{S,-} + \nu_u = \sqrt{\mu_u}, \ \forall u\in \mathbb{U},
\\
0 \leq \nu_u \leq \sqrt{\mu_u} - \sqrt{\rho_u}, \ \forall u\in \mathbb{U}.
\end{array}
\end{equation}
From equations (\ref{eq:eq-17}), it is obvious that when $ \nu_u = 0$ then $V_u^{S,+} + V_u^{S,-}= \sqrt{\mu_u}$. On the other hand,  when $ \nu_u = \sqrt{\mu_u} - \sqrt{\rho_u}$ then $V_u^{S,+} + V_u^{S,-} = \sqrt{\rho_u}$. Thus, the optimization model (\ref{eq:eq-13}) can be finally rewritten as
\begin{equation}
\label{eq:eq-18}
\begin{array}{l}
\underset{\boldsymbol{f}_{a,u}[n], \forall u,a}{\min} ~ z 
\\
\textrm{s.t.} ~
\\
 \displaystyle \sum_{i \in\mathbb{U}\setminus u} \displaystyle  \left | \sum_{a \in\mathbb{A}_i}\boldsymbol{h}_{u,a}^*[n]\boldsymbol{f}_{a,i}[n]\right |-z \leq 0, \ \forall u\in \mathbb{U},
\\ 
 V_u^{S,+} + V_u^{S,-} + \nu_u = \sqrt{\mu_u}, \ \forall u\in \mathbb{U},
\\
 0 \leq \nu_u \leq \sqrt{\mu_u} - \sqrt{\rho_u}, \ \forall u\in \mathbb{U},
\\
\displaystyle \sum_{a \in\mathbb{A}_u}\boldsymbol{h}_{u,a}^*[n]\boldsymbol{f}_{a,u}[n] - V_u^{S,+} + V_u^{S,-} = 0, \ \forall u\in \mathbb{U},
\\
 V_u^{S,+} \leq \Delta \varphi_u, \ \forall u\in \mathbb{U},
\\
 V_u^{S,-} \leq \Delta (1-\varphi_u), \ \forall u\in \mathbb{U},
\\
 V_u^{S,+} \geq 0, \ \forall u\in \mathbb{U},
\\ 
 V_u^{S,-} \geq 0, \ \forall u\in \mathbb{U},
\\
 \varphi_u \in \{0,1\}, \ \forall u\in \mathbb{U},
\\
 \displaystyle \sum_{u \in \mathbb{U}_a}\left \| \boldsymbol{f}_{a,u}[n]\right \|^2 \leq \eta_a, \ \forall a \in \mathbb{A}. 
\end{array}
\end{equation}
The formulation of the proposed optimization model in (\ref{eq:eq-18}) is tackled jointly with respect to  all variables by the cvx optimization toolbox \cite{cvx} by a single call to the MOSEK solver \cite{mosek}. 

\section{Numerical Results} \label{sec:results}
In the simulation, we assume a service area which is a square of side length 1 km and wrapped around at the boundaries. The CFmMIMO network consists of $N_A = 100$ APs uniformly distributed within the service area and each has a height of 10 m. Each AP is equipped with $M_{AP}=4$ antennas and has a maximum power budget of $\eta_a = 200 mW$. The channel bandwidth $B=20$ MHz and the ratio between the data transmission phase over the complete channel coherence block is taken equal to $\frac{\tau_d}{\tau_c} = 0.42$. We assume that 40 users are uniformly distributed within the service area where each user's device has a height of 1.65 m. The receiver's thermal noise power spectral density, $N_0$, is assumed to be $N_0=-174$ dBm/Hz, and the receiver's noise figure, $N_f$ is 9 dB. Thus, the noise variance, $\sigma_w^2=B10^{(0.1(N_f+N_0)-3)}$. We consider that every user $u$ can be served by the 15 APs that has the best large scale fading to user $u$, i.e.,  $N_{A,u}=15$. We assume an urban environment where the path from a user to an AP does not have any line of sight (LOS) components, i.e., Rayleigh small scale fading is assumed. The large scale fading coefficients, $\beta_{u,a}$ in dB, is simulated as described in \cite{8952782, 3gpp2017further}
\begin{equation}
\label{eq:eq-19}
\begin{array}{l}
\beta_{u,a}=-36.7\log_{10}(d_{u,a})-22.7\log_{10}(f)+\zeta_{u,a},
\end{array}
\end{equation}
where $d_{u,a}$ is the distance between user $u$ and AP $a$, $f$ is the operating frequency in GHz (which is taken by 1.9 GHz in our simulation), and $\zeta_{u,a} \sim \mathcal{N}(0,4^2)$ models the shadowing effect. The correlation between shadowing effects of users $u$ and $i$ is represented as in \cite{8845768} 
\begin{equation}
\label{eq:eq-20}
\mathbb{E} \{\beta_{u,a},\beta_{i,b}\}=\left\{\begin{matrix}
4^2 2^{\frac{-r_{u,i}}{r_0}}, a=b,\\ 
0, \ \ \  \ \ \ \ \ a\neq b,
\end{matrix}\right.
\end{equation}
where $r_{u,i}$ is the distance between users $u$ and $i$ and $r_0=9 m$.

In our simulation, we evaluate the performance of the proposed optimization model by benchmarking it against the conjugate beamforming (CBF) with both uniform power allocation (UPA) and proportional power allocation (PPA). In CBF-UPA, the beamforming vectors are given by $\boldsymbol{c}_{a,u}^{UPA}[n]=\sqrt{\frac{\eta_a}{N_{U,a}}} \frac{\boldsymbol{h}_{u,a}}{\left \| \boldsymbol{h}_{u,a} \right \|}, \forall a,u$. In the CBF-PPA, the beamforming vectors are $\boldsymbol{c}_{a,u}^{UPA}[n]=\sqrt{\frac{\eta_a}{\sum_{u \in \mathbb{U}_a}\left \| \boldsymbol{h}_{u,a} \right \|^2}} \boldsymbol{h}_{u,a}, \forall a,u$. We perform simulation for 250 iterations and evaluate each of the considered algorithms in terms of the individual users data rates, the individual APs consumed power and the total system's DL radio energy efficiency (EE). For the proposed scheme, we use $\Delta =1000$. Also, we set the lower bound of the desired signal $\rho_u$ to be
\begin{equation}
\label{eq:eq-21}
\begin{array}{l}
\rho_u=\sigma_w^2+\frac{\eta_a}{N_AN_U}\left \| \boldsymbol{h}_{u,t} \right \|^2,
\end{array}
\end{equation}
where $t$ identifies the AP in the set $\mathbb{A}_u$ that has the best channel to user $u$. Furthermore, the upper bound of the desired signal $\mu_u$ is set to be 
\begin{equation}
\label{eq:eq-22}
\begin{array}{l}
\mu_u=10^5 \sigma_w^2+\frac{\eta_a\sqrt{N_A}}{N_U}\left \| \boldsymbol{h}_{u,t} \right \|^2.
\end{array}
\end{equation}

\begin{figure}[!t]
\begin{center}
\includegraphics[width =0.85\columnwidth]{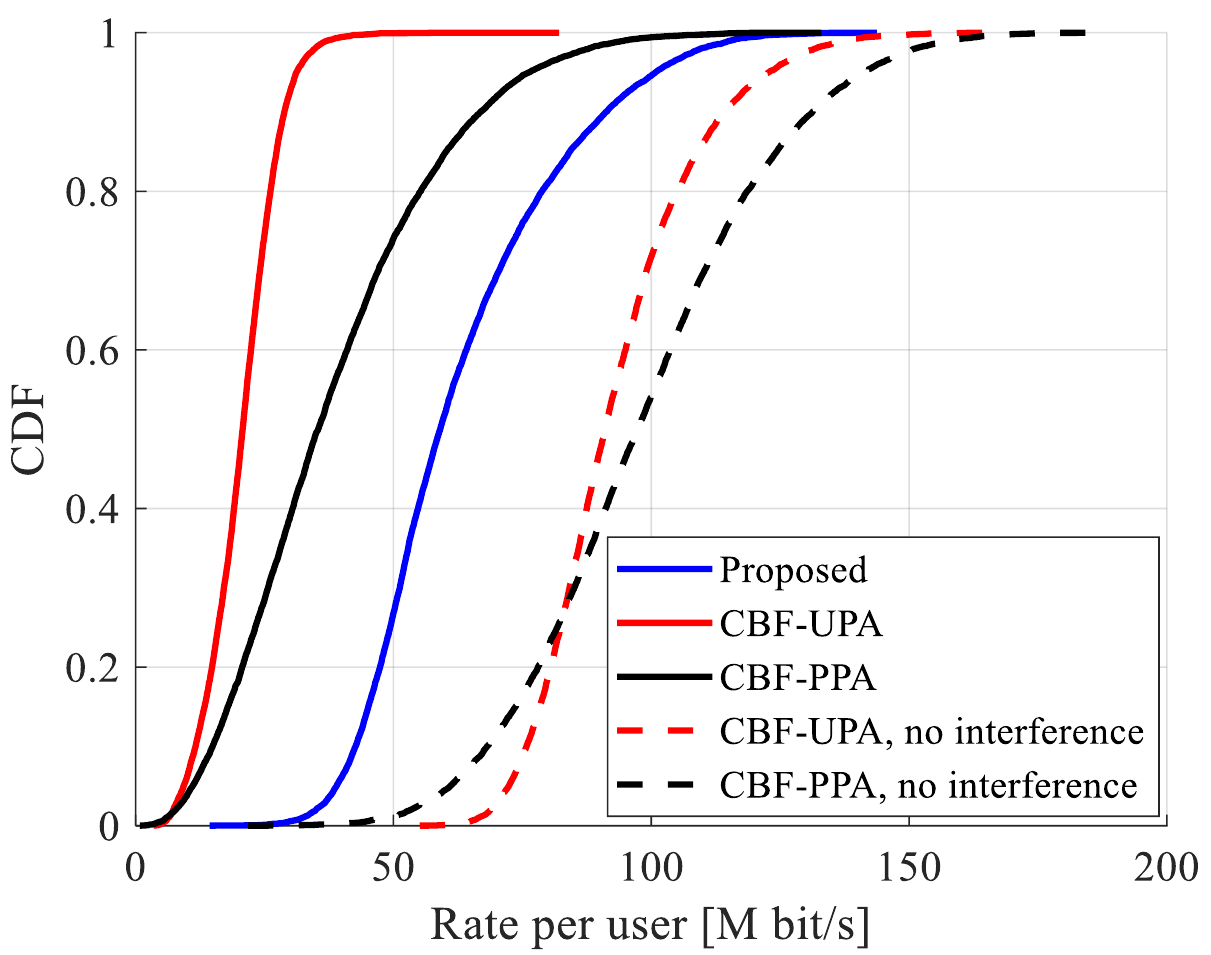}
\caption{Individual users data rate comparison among the proposed joint beamforming and power control scheme, the CBF-UPA, the CBF-PPA in a CFmMIMO system.}
\label{fig:Rate_all_users}
\end{center}
\end{figure}
Figure \ref{fig:Rate_all_users} shows the cumulative distribution function (CDF) plots of the individual users' DL rate for the considered algorithms, where all users' rates in every iteration are taken into account. It is clear that the proposed joint beamforming and power control algorithm outperforms both CBF-UPA and CBF-PPA in terms of the individual users' data rates. 
Besides, the 95\% likely users' rate of the proposed scheme is about 4.33 times and 3.55 times the counterparts of CBF-UPA and CBF-PPA, respectively. In this figure, we also show the unachievable no-interference upper bounds for both CBF-UPA and CBF-PPA (plotted in dashed lines) to get more insights about how much the proposed scheme improves the data rate performance from existing solutions in the way towards their bounds. 

\begin{figure}[!t]
\begin{center}
\includegraphics[width =0.85\columnwidth]{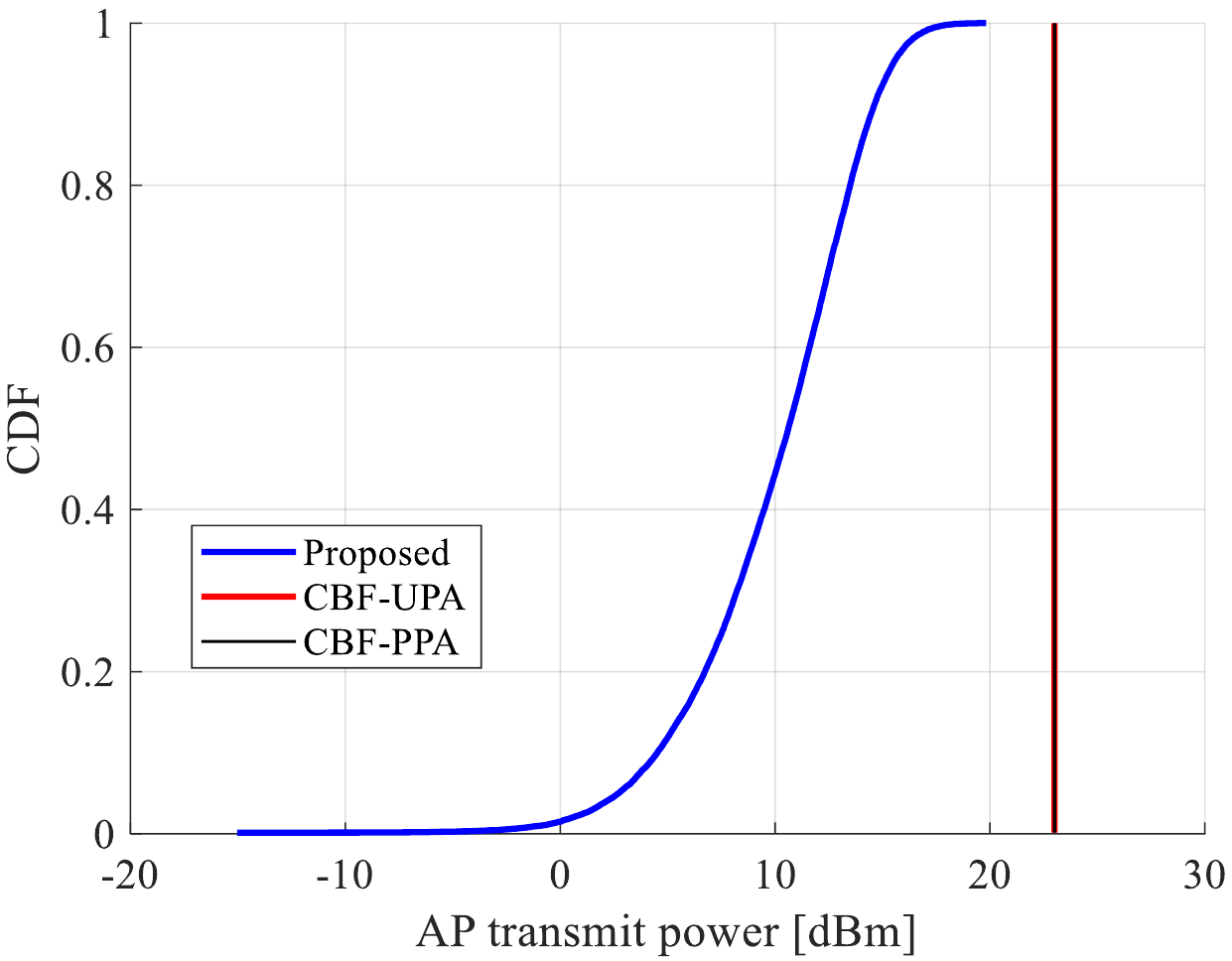}
\caption{Individual APs transmission power comparison among the proposed joint beamforming and power control scheme, the CBF-UPA, the CBF-PPA in a CFmMIMO system.}
\label{fig:Power_all_APs}
\end{center}
\end{figure}

The CDF plots of the individual APs' transmission power for the considered schemes are depicted in Fig. \ref{fig:Power_all_APs}. It is noticeable that almost always APs are transmitting at their full power in both CBF-UPA and CBF-PPA. On the other hand, the proposed scheme has more flexibility, where the AP transmission power can be reduced as long as the required bounds on the desired signal are satisfied. In the case of the proposed scheme, the worst AP's transmit power is less than that of its counterparts of CBF-UPA and CBF-PPA by at least 3.19 dB. 

\begin{figure}[!t]
\begin{center}
\includegraphics[width =0.85\columnwidth]{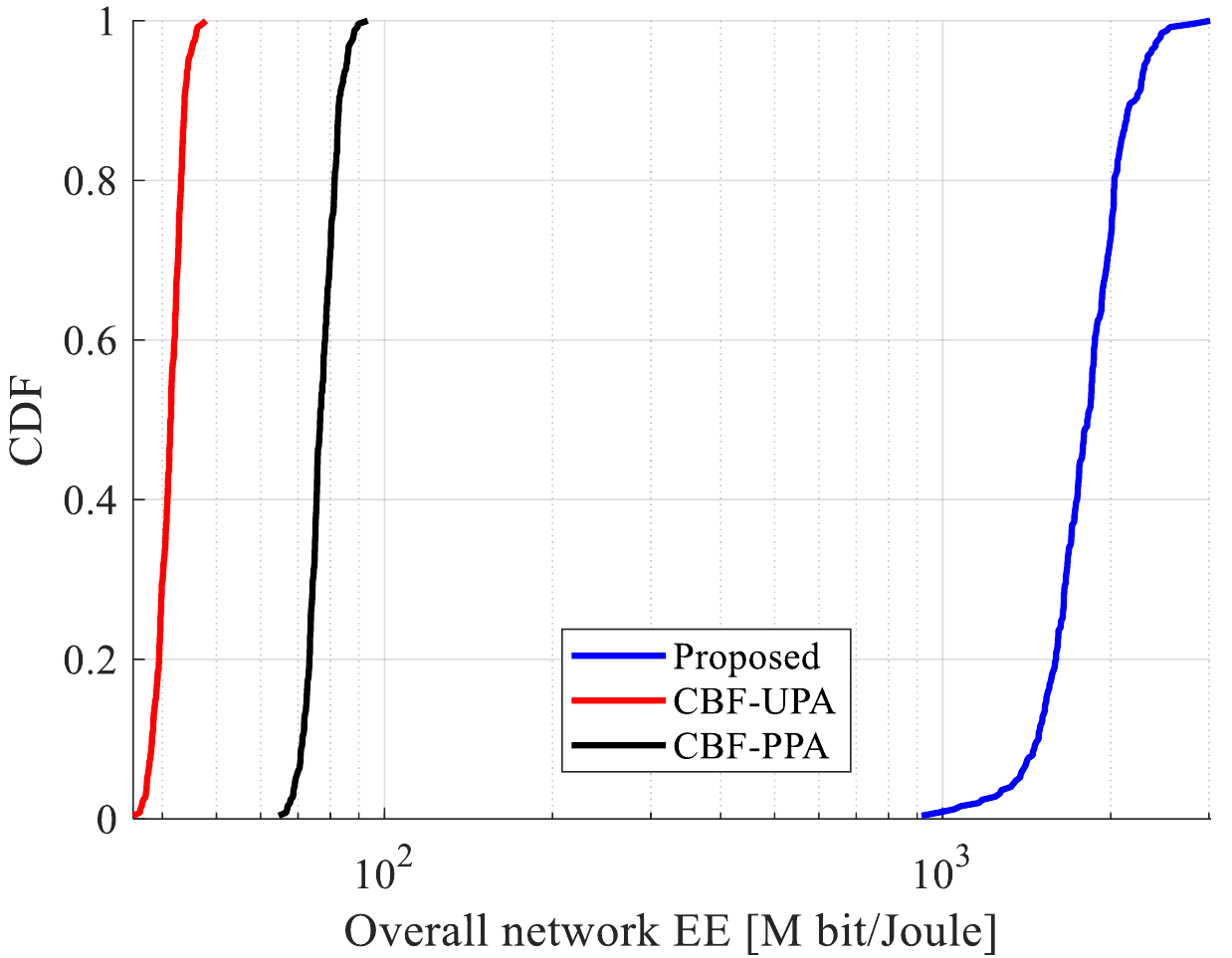}
\caption{Overall network radio energy efficiency comparison among the proposed joint beamforming and power control scheme, the CBF-UPA, the CBF-PPA in a CFmMIMO system.}
\label{fig:Overall_EE}
\end{center}
\end{figure}

Figure \ref{fig:Overall_EE} depicts the CDF plots of the overall DL radio EE for the considered schemes. Here, the radio EE is the ratio of the overall DL sum data rate to the total sum of APs' transmission power. Fig. \ref{fig:Overall_EE} clearly highlights that the proposed joint beamforming and power control scheme provides significant improvements in the overall DL radio EE compared to both CBF-UPA and CBF-PPA. Specifically, the least overall DL radio EE of the proposed joint beamforming and power control scheme is about 19.16 times the best overall DL radio EE of the CBF-UPA. Moreover, the least overall DL radio EE of the proposed scheme is about 9.82 times the best overall DL radio EE of the CBF-PPA. Regarding the computational complexity of the proposed optimization model, our numerical results indicates that the proposed model can be solved for the considered simulation parameters by the MOSEK solver under the cvx toolbox within about 1 minute on a 3.6 GHz Intel  processor.

\section{Conclusion} \label{sec:conclusion}
This paper has presented a novel optimization framework for joint beamforming and power control in the CFmMIMO network. The proposed model is based on minimizing the worst user's interference with upper and lower bounds on the desired signal power at the user along with constraints on the AP transmission power budget. The proposed optimization model is transformed to a form that is directly tractable by using cvx optimization toolbox. Our simulation results prove that the proposed scheme improves the individual users' data rates, reduces the APs transmission power, and boosts the network radio energy efficiency compared to CBF-UPA and CBF-PPA.

\bibliographystyle{IEEEtran}
\bibliography{bibfile}
\end{document}